# Fledgling quantum spin Hall effect in pseudo gap phase of Bi2212

*Udai Prakash Tyagi, Kakoli Bera, and Partha Goswami(Retd.)\**

*Deshbandhu College, University of Delhi, Kalkaji, New Delhi-110019,India*

*\*E-mail of the corresponding author: physicsgoswami@gmail.com; E-mail of the first author: uptyagi@yahoo.co.in, E-mail of the second author:kakolibera@gmail.com;*

**ABSTRACT**

We study the emergence of the quantum spin Hall (QSH) states for the pseudo-gap (PG) phase of Bi2212 bilayer system, assumed to be D-density wave(DDW) ordered, starting with a strong Rashba spin-orbit coupling(SOC) armed, and the time reversal symmetry (TRS) complaint Bloch Hamiltonian. The presence of strong SOC gives rise to non-trivial, spin-momentum locked spin texture tunable by electric field. The emergence of quantum anomalous Hall effect with TRS broken Chiral DDW Hamiltonian of Das Sarma et al. is found to be possible.

**1.Introduction**

We consider two model Bloch Hamiltonians for the pseudo-gap (PG) phase of $Bi_2Sr_2CaCu_2O_{8+\delta}$ (Bi2212) bilayer system where the first model is time reversal symmetry (TRS) protected and the second one is non-protected. The strong Rashba spin-orbit coupling(RSOC) is present in both the cases and therefore the Hamiltonians are inversion asymmetric. The former corresponds to $d_{x^2-y^2}$-density wave (DDW) **[1-3]**ordered phase whereas the latter to $d_{xy} + i\ d_{x^2-y^2}$ (Chiral DDW) orderded phase **[4-6]**. The models also involve a term accounting for the effect of coupling between different $CuO_2$ planes. The origin of RSOC lies in the fact that whereas one Cu-O layer has Ca ions above and, Bi-O ions below, in the unit cell of the other layer this situation is reversed. This leads to a nonzero electric field within the unit cell. We find spin-momentum locking (SML) in momentum space which suggests the presence of a strong spin-orbit coupling in Bi2212 like a topological insulator. A recent spin- and angle-resolved photoemission spectroscopy measurement for $Bi_2Sr_2CaCu_2O_{8+\delta}$ had reported[7] a non-trivial spin texture which corresponds to a well-defined direction for each electron real spin depending on its momentum.

The focal point of the communication is to investigate the possibility of occurrence of the quantum spin Hall(QSH) effect when the time reversal symmetry is protected.In general, upon having two quantum Hall insulating layers, one over the other, leads to a trivial insulator. However, our band structure based detailed analysis has indicated that Bi2212 will have access to QSH**[8]** phase provided we have strong SOC and invariance of TRS. WhenTRS is broken in a 2D system, say due to the presence of the magnetic impurities (but no magnetic field), the system spontaneously crosses over to QAH phase at low temperatures **[9]**. A quantum transition in the transverse Hall resistance, accompanied by a considerable drop in longitudinal resistance, is then expected. This is a prominent indicator of QAH. The anomalous effect is an intrinsic quantum-mechanical property of a perfect crystal as it involves the Berry-phase curvatures.

The pseudo-gap is a distinct phase akin to an unconventional metal or, a symmetry preserved/broken state [1-6, 10-22]. As there is no consensus yet on this issue, the exact nature of the pseudo-gap still remains an enigma. The onset of the pseudo-gap is defined by the opening of an anti-nodal gap and the reduction of the large Fermi surface to Fermi arc [1-6, 10-22] as in Wyle/Dirac semi-metal. As mentioned above, n this communication, the PG state has been first identified with the DDW (d + i0)state and subsequently with the CDDW (d+id) state. The former preserves the TRS, whereas in the case of the latter the imaginary part of the d+id-wave order parameter $D_k=(-\chi_k+i\Delta_k)$ breaks the time reversal symmetry. Thus, whereas CDDW Hamiltonian is suitable to investigate the QAH state, the DDW hamiltonian is suitable for QSH. The quantities ($\chi_k$, $\Delta_k$) are given by [4-6] $\chi_k = -(\chi_0/2)$ $\sin(k_x a)\sin(k_y a)$, $\Delta_k = (\Delta_0^{(PG)}(T)/2)(\cos k_x a - \cos k_y a)$, and $\mathbf{k} = (k_x, k_y)$. The ordering wave vector $\mathbf{Q}$ is taken to be $\mathbf{Q}=(Q_1=\pm\pi, Q_2=\pm\pi)$. We mention that the the onset of CDDW ordering leads to a peak in the anomalous Nernst signal (ANS)[4-6]. The main contribution to this chirality induced ANS comes from the points $(\pm\pi(1-\varphi), \pm\pi\varphi),(\pm\pi\varphi, \pm\pi(1-\varphi))$ with $\varphi \sim 0.2258$ located roughly on the boundary of the Fermi pockets in the momentum space( cf. ref. [4,6]).All energies in our calculation below are expressed in units of the first neighbor hopping. The second-neighbor hopping in the dispersion, which is known to be important for cuprates [1-6, 20-22], frustrates the kinetic energy of electrons.

The paper is organized as follows: In section 2 we derive an expression for the single-particle excitation spectrum in PG state and discuss QSH effect. In section 3, we carry out an investigation of the QAH effect. The paper ends with brief discussion and conclusion in section 4.

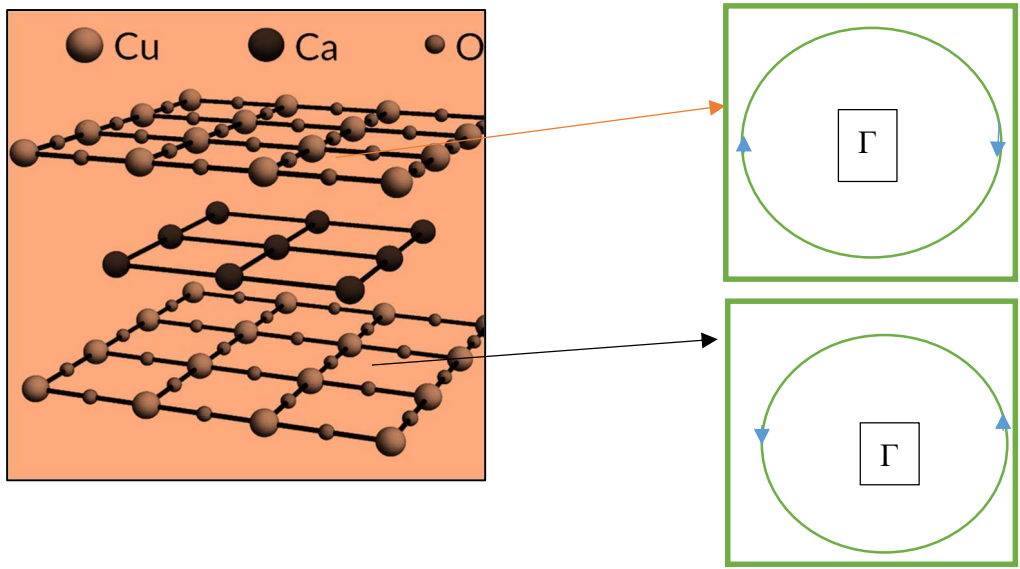

**Figure 1.** A structural unit of Bi2212 comprising of two structurally equivalent $CuO_2$ planes separated by calcium (Ca) layer. On the right hand side, we have the momentum space spin pattern corresponding to two adjacent CuO$_2$ layers within a unit cell in ref.[7]. The Γ point is encircled by oppositely pointed spins in the two representations.

## 2. Quantum Spin Hall Insulator

Recent studies have demonstrated that the spin-polarized carriers injected from a ferromagnet (F) into a nonmagnetic material (N), such as a normal conducting metal, semiconductor, and superconductor, give rise to spin accumulation and spin current over the diffusion length. Thus, electrical currents generate (transverse) spin currents. This phenomenon, linked to relativistic spin-orbit coupling, is referred to as the spin Hall effects (SHE).A spin version of quantum Hall effect, on the other hand, is called QSH effect. A QSH system possesses a twisted Hilbert space leading to a pair of helical edge states with opposite spins propagating in opposite directions. Here an electric current can induce a transverse spin current near the system boundary. Such non-trivial states with spin-momentum locking gives rise to two-dimensional topological insulator. The chiral edge states around boundary in a QAH system give rise to quantized Hall conductance even in the absence of external magnetic field. In what follows we show that the Hilbert space of our system Bi2212 bilayer is twisted. The physical consequence of this nontriviality, as already stated, is the appearance of topologically-protected edge states.

The cuprate Bi2212 consists of CuO$_2$ layers separated by spacer layers as in Bi2212 as shown in Figure 1. The total dispersion for non-interacting system, together with the effect of coupling between different CuO$_2$ planes, can be written as $\epsilon(k,k_z) = \varepsilon_k + \varepsilon_{k_z}(k)$, where $\varepsilon_k = e_k - \mu$, **k** and $k_z$ respectively denote the in-plane and out-of-plane components of **K = (k, $k_z$)**. The term μ stands for the chemical potential of the fermion number. The dispersion $e_k = -2t\, P_1(k) + 4t_1 P_2(k) - 2t_2 P_3(k) - 4t_3 P_4(k)$ including the in-plane first, the second, the third neighbor, and the fourth neighbor hoppings. Here $P_1(k) = \left(s_x(a) + s_y(a)\right), P_2(k) = s_x(a)\, s_y(a),\ P_3(k) = \left(s_x(2a) + s_y(2a)\right),$ and $P_4(k) = \left(s_x(a)s_y(2a) + s_x(2a)s_y(a)\right)$. For the hole-doped (electron-doped) materials, $t_1 > 0$ ($t_1 < 0$), and, in all cases, $t_1 <$ (t/2). The $\varepsilon_{k_z}(k)$ term accounts for the effect of coupling between different CuO$_2$ planes, and possesses the following form for bilayer Bi2212 [23]: $\varepsilon_{k_z}(k) = -\Upsilon_z(k, s_z(c/2))[(s_x(a) - s_y(a))^2/4 + a_0]$ where c denotes the lattice constant along the z-axis, and the term $\Upsilon_z(k, s_z(c/2)) = \pm\, [t_p^2 + 16\, t_{in}^2 s_x^2\left(\frac{a}{2}\right) s_y^2\left(\frac{a}{2}\right) + 8t_p t_{in} s_x\left(\frac{a}{2}\right) s_y\left(\frac{a}{2}\right) s_z\left(\frac{c}{2}\right)]^{1/2}$. Here $t_p$ is an effective parameter for hopping within a single bilayer, i.e. it controls the intracell bilayer splitting and $s_j(a) \equiv \cos(k_j a)$ [24]. The compound Bi2212, however, involves intercell hopping ($t_{in}$), too. The couplings $t_p$ and $t_{in}$ and the in-plane hopping terms beyond the NN term are small compared to the NN hopping. The plus (minus) sign refers to the bonding (anti-bonding) solution. The term $c_z$ arises because supposedly we have an infinite number of stacked layers. Along the high symmetry line X($\pi$,0) —M($\pi$, $\pi$), we have $\Upsilon_z$ = ± $t_p$. This leads to a lack of $k_z$-dispersion along this high-symmetry line, for $\varepsilon_{k_z}(k) \to \varepsilon(k_y) = \pm(1/4)\ t_p\ [(1+s_y(a))^2 + 4a_0]$. The additional hopping $a_0$ allows for the presence of a splitting at $\Gamma$(0,0). It is reported [25] that adequate control of the interlayer spacing albeit the interlayer hopping in Bi-based superconductors is possible through the intercalation of guest molecules between the layers. This could be a way to tune the hopping parameter.

As already stated, motivated by the findings of S. Das Sarma et al.[4-6], we assume that the CDDW order represents the pseudo-gap(PG) phase. We assume the momentum dependence of the pairing interactions required for this kind of ordering is given by the functions of the form $U_{x2-y2}(k,k')$ = $U_1$ ($cosk_xa$–$cosk_ya$)($cosk'_xa$–$cosk'_ya$), $U_{xy}(k,k')$= $U_2$sin($k_xa$) sin($k_ya$) sin($k'_x a$) sin($k'_ya$), where $U_1$ and $U_2$ ($U_1$< $U_2$ ) are the coupling strengths, and ($k_x,k_y$) belong to the first Brillouin zone (BZ). These two interactions are required for QAH state. Our quest for QSH state, however, demands that we need to consider only the interaction $U_{x2-y2}(k,k')$ = $U_1$ ($cosk_xa$–$cosk_ya$)($cosk'_xa$–$cosk'_ya$). Consequently, the PG state of Bi2212 will be a d-density wave state of Chakravorty et al.[1-3]. If we further assume that the order parameter in superconducting state has d-wave symmetry as is demanded by the spin-fermion model, we may sound like advocates of pre-formed Cooper pair formation [ 26,27]. Also, motivated by the findings of Vishwanath et al.[7] we introduce the Rashba-coupling as a special ingradient. Suppose now $d_{k,\sigma}$ ( $\sigma$ = ↑↓ for the real spin ) corresponds to the fermion annihilation operator for the single-particle state $\{\mathbf{k} = (k_x, k_y), \sigma\}$ in a single layer of the system. In the basis $(d^\dagger_{k,\uparrow}\ d^\dagger_{k,\downarrow}\ d^\dagger_{k+Q,\uparrow}\ d^\dagger_{k+Q,\downarrow})^T$, we consider a 2 $D$ mom − entum space Hamiltonian

$$H(k_x,k_y) = \begin{pmatrix} H_0(k_x,k_y) & \varsigma(k_x,k_y) \\ \varsigma^*(k_x,k_y) & H_0(k_x+Q_1,k_y+Q_2) \end{pmatrix} \qquad (1)$$

to describe the pseudo-gap phase of the system involving CDDW order given by the matrix $\varsigma(k_x,k_y) = \begin{pmatrix} D^\dagger_k & 0 \\ 0 & D_k \end{pmatrix}$, where $D_k = (-\chi_k + i\Delta_k)$ breaks the parity and the time reversal symmetry of the normal state. The matrix $H_0(k,\mu) = \begin{pmatrix} \varepsilon_k & \alpha(k_x,k_y) \\ \alpha^*(k_x,k_y) & \varepsilon_k \end{pmatrix}$, where the intra-layer Rashba spin-orbit coupling $\alpha(k_x,k_y) = \alpha_k = \alpha_0(-i sin(k_xa) - sin(k_ya))$. The quantity $\alpha_0$ is the RSO coupling strength which is proportional to the strength of the nonzero electric field within the unit cell ( mentioned in section 1) assumed to be in the z direction. As regards the e-e interactions, $U\ \sum_i\ d^\dagger_{i\uparrow}d_{i\uparrow}d^\dagger_{i\downarrow}d_{i\downarrow}$ is the onsite repulsion of $d$ electrons, where the intra-layer $d$ electrons are locally interacting via a Hubbard-$U$ repulsion. We have not considered this term assuming the correlation effect are marginally relevant. As noted above, the Hamiltonian (1) is suitable to investigate the QAH state. For the QSH state, we have $D_k = \Delta_k = (\Delta_0^{(PG)}(T)/2)(\cos k_x a - \cos k_y a)$, which breaks the parity only. The investigation is expected to show spin-momentum locking in some regions of BZ.

The eigenvalues $\varepsilon$ derivable from the Hamiltonian (1), suitable to investigate QAH effect, are given by the quartic $\varepsilon^4 + a_A(k)\varepsilon^3 + b_A(k)\varepsilon^2 + c_A(k)\varepsilon + d_A(k) = 0$. In view of the Ferrari's solution of a quartic equation, we find the roots as

$$\epsilon_j(s,\sigma,k) = \sigma\sqrt{\frac{\eta_0(k)}{2}} + \varepsilon_k^U + s\left(b_0(k) - \left(\frac{\eta_0(k)}{2}\right) + \sigma c_0(k)\sqrt{\frac{2}{\eta_0(k)}}\right)^{\frac{1}{2}}, \quad (2)$$

where j= 1,2,3,4, $\sigma = \pm 1$ is the spin index and $s = \pm 1$ is the band-index. Since, the spin index $\sigma$ occurs twice in Eq. (2), the term $\sqrt{\frac{z_0(k)}{2}}$ does not act like magnetic energy. The coefficients are $a_A(k) = -4\varepsilon_k^U$, $b_A(k) = \{4\varepsilon_k^{U^2} + 2(\varepsilon_k \varepsilon_{k+Q} - |D_k|^2) - |\alpha_k|^2 - |\alpha_{k+Q}|^2\}$, $c_A(k) = -4\varepsilon_k^U(\varepsilon_k \varepsilon_{k+Q} - |D_k|^2) - 2\varepsilon_{k+Q}|\alpha_k|^2 - 2\varepsilon_k|\alpha_{k+Q}|^2$, $d_A(k) = (\varepsilon_k \varepsilon_{k+Q} - |D_k|^2)^2 - \varepsilon_{k+Q}^2|\alpha_k|^2 - \varepsilon_k^2|\alpha_{k+Q}|^2 + |\alpha_k|^2|\alpha_{k+Q}|^2 + F(k_x, k_y)$, where $\varepsilon_k^U = (\varepsilon_k + \varepsilon_{k+Q})/2$, $|D_k|^2 = (\chi_k^2 + \Delta_k^2)$, $\alpha_k = \alpha_0(-i\sin(k_x a) - \sin(k_y a))$ and $F(k_x, k_y) = -2\alpha_0^2(\chi_k^2 - \Delta_k^2)f(k_x, k_y) - 4\alpha_0^2(\chi_k \Delta_k)g(k_x, k_y)$, $f(k_x, k_y) = [\sin(k_x a)\sin(k_x a + Q_1) + \sin(k_y a)\sin(k_y a + Q_2)]$, and $g(k_x, k_y) = [\sin(k_x a)\sin(k_y a + Q_2) - \sin(k_y a)\sin(k_x a + Q_1)]$. The functions appearing in (2) are given by

$$\eta_0(k) = \frac{2b_0(k)}{3} + (\Delta(k) - \Delta_0(k))^{\frac{1}{3}} - (\Delta(k) + \Delta_0(k))^{\frac{1}{3}},$$

$$\Delta_0(k) = \left(\frac{b_0^3(k)}{27} - \frac{b_0(k)d_0(k)}{3} - c_0^2(k)\right),$$

$$\Delta(k) = \left(\frac{2}{729}b_0^6 + \frac{4d_0^2 b_0^2}{27} + c_0^4 - \frac{d_0 b_0^4}{81} - \frac{2b_0^3}{27} + \frac{2c_0^2 b_0 d_0}{3} + \frac{d_0^3}{27}\right)^{1/2},$$

$$b_0(k) = \left\{\frac{3a_A^2(k) - 8b_A(k)}{16}\right\}, \quad c_0(k) = \left\{\frac{-a_A^3(k) + 4a_A(k)b_A(k) - 8c_A(k)}{32}\right\}$$

$$d_0(k) = \frac{-3a_A^4(k) + 256d_A(k) - 64a_A(k)c_A(k) + 16a_A^2(k)b_A(k)}{256}. \quad (3)$$

In order to investigate QSH effect, we need to have invariant TRS. Therefore, we have to make the following substitutions: $\chi_k = 0$, $|D_k|^2 = \Delta_k^2$, $F(k_x, k_y) = 2\alpha_0^2 \Delta_k^2 [\sin(k_x a + Q_1)\sin(k_x a) + \sin(k_y a)\sin(k_y a + Q_2)]$, and $\mathbf{Q}=(Q_1=\pm\pi, Q_2=\pm\pi)$. For the bilayer system, the energy eigenvalues are given by

$$E_j(s,\sigma,k,k_z) = \epsilon_j(s,\sigma,k) \pm \varepsilon_{k_z}(k) \quad (4)$$

for the bonding as well as the anti-bonding cases. It must be mentioned that an anisotropic 3D Rashba-like spin splitting has been directly observed by spin-ARPES **[28]**, which gave rise to emergence of large spin splitting at the M points of the Brillouin zone instead of the Γ point, where Rashba type splitting is usually found. We have not considered this interlayer RSOC here assuming it to be small compared to the other terms in $E_j(s,\sigma,k,k_z)$.

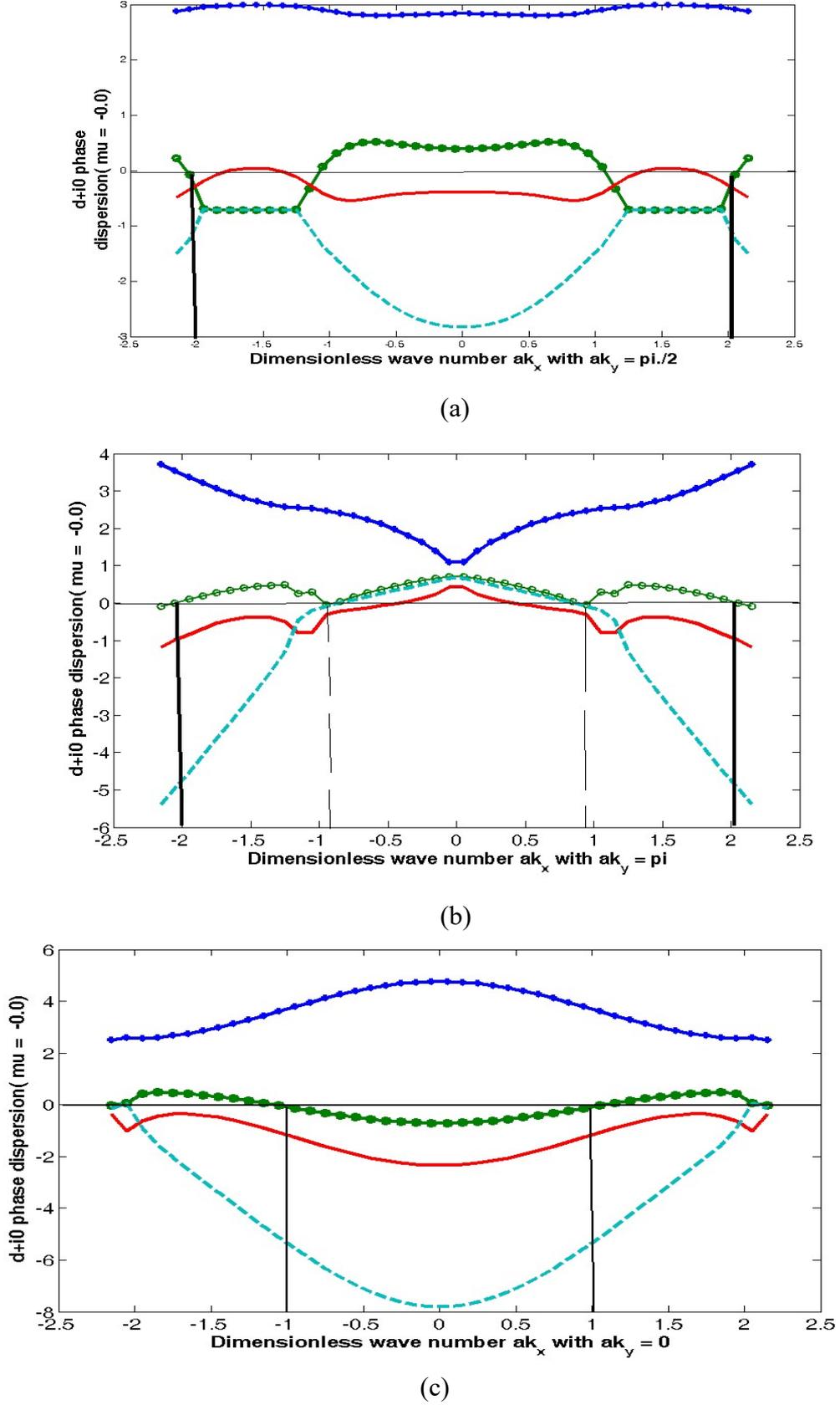

(a)

(b)

(c)

**Figure 2.** The energy spectra of electron dispersion $\epsilon_j (s, \sigma, k)$ given by (2) in DDW state (QSH phase) without the interlayer tunneling at (a)the nodal($ak_x \sim \frac{\pi}{2}, ak_y = \frac{\pi}{2}$)region and (b) the anti-nodal( $ak_x \sim 0, ak_y = \pi$)region, and Γpoint. The chemical potential μ is represented by solid, horizontal line μ ~0 is located as shown. In figures (a), (b), and (b), the common feature is the Fermi energy $E_F = 0$ inside the gap intersects the layer bands (in the layer multi-band

structures) in the same BZ an odd (one) number of times. The parameter values are, t =1 , t′/t = −0.28(hole-doping), t ″/t = 0.1, t ‴/t = 0.06, $t_p$/t = 0.3, $t_{in}$/t = 0.1, $\frac{\alpha_0}{t}$ = 0.95, μ= −0.0, and $\Delta_0^{(PG)}$ (T)= 0.02. The state dispersion is perfectly nested with the ordering wave vector **Q**=($Q_1$=±π, $Q_2$= ±π). The red and blue bands highlight the two bands involved in band inversion. The vertical lines are the indicators of time reversal invariant momenta(TRIM) where $ak_{trim} = (2, \frac{\pi}{2})$, (1,0) in (a), (b) and (c).

The plots of energy bands $\epsilon_j (s, \sigma, k)$ in (2) in the d+i0 (QSH) case belonging to **the nodal and anti-nodal regions**, as well as at Γ point are shown in Figures 2. The red and blue bands highlight the two bands involved in band inversion. The numerical values of the parameters to be used in the calculation are t =1 , t′/t = −0.28(hole-doping), t ″/t = 0.1, t ‴/t = 0.06, $t_p$/t = 0.3, $t_{in}$/t = 0.1, $\frac{\alpha_0}{t}$ = 0.95, μ= −0.0, and $\Delta_0^{(PG)}$ (T)= 0.02. The state dispersion is perfectly nested with the ordering wave vector ***Q***=($Q_1$=±π, $Q_2$= ±π). Throughout the whole paper, we choose ***t*** to be the unit of energy. On a quick sidenote we observe that the Fermi energy $E_F$ inside the gap intersects the layer bands (in the layer multi-band structures) in the same BZ either an even or odd number of times. If there are odd number of intersections at TRIMs, which guarantees the time reversal invariance, the layer state is topologically non-trivial (strong topological insulator), for disorder or correlations cannot remove pairs of such surface state crossings (SSC) by pushing the surface bands entirely above or below the Fermi energy $E_F$. This has been checked by assigning different values to the chemical potential. We notice that the number of paired SSCs in each of the figures in Figure 2 is one(odd). We also notice that $ak_{trim} = (2, \frac{\pi}{2})$, (1,0) in (a), (b) and (c). The vertical lines are the indicators of time reversal invariant momenta(TRIM). Amongst the pair of TRI momentum (**K,−K),** the relation $-\mathbf{K} = \mathbf{K} + \mathbf{G}$ is satisfied. When there are even number of pairs of layer-state crossings, the surface states are topologically trivial (weak TI or conventional insulator), for disorder or correlations can remove pairs of such crossings. The inescapable conclusion is that the system under consideration is quantum spin Hall insulator, or a strong topological insulator. The Bi2212 layer, therefore, comprises of 'helical liquids'[29] in the monolayer (2D) geometry, which (helicity) is one of the most unique properties of a topologically protected surface. These lead to the conclusion that the system considered is a strong TI or a QSH material. A recent report of a spin-signal on the surface by the inverse Edelstein effect **[30]** has confirmed the helical spin-structure.

We now calculate below topological invariant ν to characterize the present 2D quasiparticle systems preserving TRS in the same manner as formulated by Fu and Kane **[31]**. We require the eigenvectors corresponding to the energy eigenvalues $\epsilon_j (s, \sigma, k)$. These are Bloch states given by

$$|u^{(\alpha)}(k)\rangle = \varsigma_\alpha^{-1/2}(k) \begin{pmatrix} \psi_1^\alpha(k) \\ \psi_2^\alpha(k) \\ \psi_3^\alpha(k) \\ \psi_4^\alpha(k) \end{pmatrix}, \quad \alpha = 1, 2, 3, 4,$$

$$\zeta_\alpha(k) = |\psi_1^\alpha(k)|^2 + |\psi_2^\alpha(k)|^2 + |\psi_3^\alpha(k)|^2 + |\psi_4^\alpha(k)|^2$$

$$\psi_1^\alpha(k) = 1, \quad \psi_\eta^\alpha(k) = \Delta_\eta^{(\alpha)}(k)/\Delta^{(\alpha)}(k), \eta = 2,3,4$$

$$\Delta^{(j)}(k) = \alpha_k D_k(\in_j(s,\sigma,k) - \varepsilon_{k+Q}) + \alpha_{k+Q} D_k^\dagger(\in_j(s,\sigma,k) - \varepsilon_k),$$

$$\Delta_2^{(j)}(k) = D_k(\in_j(s,\sigma,k) - \varepsilon_k)(\in_j(s,\sigma,k) - \varepsilon_{k+Q}) + D_k^\dagger \alpha_k^* \alpha_{k+Q} - D_k |D_k|^2,$$

$$\Delta_3^{(j)}(k) = -\alpha_{k+Q}|\alpha_k|^2 + \alpha_k D_k^2 + (\in_j(s,\sigma,k) - \varepsilon_k)^2 \alpha_{k+Q},$$

$$\Delta_4^{(j)}(k) = -|\alpha_k|^2(\in_j(s,\sigma,k) - \varepsilon_{k+Q}) - |D_k|^2(\in_j(s,\sigma,k) - \varepsilon_k)$$

$$+ (\in_j(s,\sigma,k) - \varepsilon_{k+Q})(\in_j(s,\sigma,k) - \varepsilon_k)^2. \quad (5)$$

The Hamiltonian H(k)(QSH) satisfies $[H(k),\Theta] = 0$. This means H(k) satisfies $\Theta^{-1} H(-k) \Theta = H(k)$. Here the time reversal (TR) operator $\Theta$ for a spin 1/2 particle takes the simple form $\Theta = I \otimes \sigma_y K$. The $\sigma_j$ are Pauli matrices on two dimensional $\mathbf{k}$ − space. The operator K is for the complex conjugation. We consider now a matrix representation of the TR operator in the Bloch wave function basis. The representation is $\vartheta_{\alpha\beta}(k) = \langle u^{(\alpha)}(-k)|\Theta|u^{(\beta)}(k)\rangle$, where α and β are band indices. With the aid of one can easily show that $\vartheta_{\alpha\beta}(k)$ is a unitary matrix. We also find that it has the property:

$$\vartheta_{\alpha\beta}(-k) = i\left[\psi_2^{*\beta}(k) - \psi_2^{*\alpha}(-k) + \psi_4^{*\beta}(k)\psi_3^{*\alpha}(-k) - \psi_3^{*\beta}(k)\psi_4^{*\alpha}(-k)\right]$$

$$=. -\vartheta_{\beta\alpha}(k). \quad (6)$$

This implies that at a TRIM $k_{trim}$ we have $\vartheta_{\alpha\beta}(k_{trim}) = -\vartheta_{\beta\alpha}(k_{trim})$, i.e. the matrix $\vartheta_{\alpha\beta}(k_{trim})$ becomes anti-symmetric.

We now consider the green and red bands in Figure 2(a), and denote their Bloch wave functions, respectively, as $|u^{(2)}(k)\rangle$ and $|(u^{(3)}(k)\rangle$. Assume that the band parameters change with time and return to the original values at t = T. Furthermore, we consider the situation when the 1D Hamiltonian satisfies the following conditions $H[t+T] = H[t]$ and $H[-t] = \Theta H[t] \Theta^{-1}$. It is well-known [31] that charge polarization P can be calculated by integrating the Berry connection of the occupied states over the BZ. In the present case of the 1D two-band system, P may be written as P = $P_2$ + $P_3$ where the Berry connections $\{-i\langle u^{(j)}(k)|\nabla_k|u^{(j)}(k)\rangle\}$ are given by $c_{jj}(k)$ (j = 2,3) and $P_2 = \int_{-\pi}^{\pi}\frac{dk}{2\pi} c_{22}(k)$, $P_3 = \int_{-\pi}^{\pi}\frac{dk}{2\pi} c_{33}(k)$. The Berry curvature is given by $\Omega_j(k) = \nabla_k \times c_{jj}(k)$. The total polarization density $C(k) = c_{22}(k) + c_{33}(k)$. This yields the TR polarization is defined by $P_{tr} = P_2 - P_3$ = $2P_2$ − P. Here $P_{tr}$ gives the difference in charge polarization between spin-up and spin-

down quasiparticle bands. We now go back to Eq.(7) which gives the charge polarization P, calculated by integrating the Berry connection of the occupied states over the BZ. Furthermore, the time-reversed version of $|(u^{(3)}(k)\rangle$, i.e. $\Theta |(u^{(3)}(k)\rangle$, is equal to $|u^{(2)}(-k)\rangle$ except for a phase factor. Hence, at t = 0 and t = T/2 one may write $\Theta |u^{(3)}(k)\rangle = e^{-i\rho(k)}|u^{(2)}(-k)\rangle$ and $\Theta |u^{(2)}(k)\rangle = -e^{-i\rho(-k)}|u^{(3)}(-k)\rangle$. It is not difficult to see that the matrix $\vartheta(k)$ will now be given as $\vartheta(k) = \begin{pmatrix} 0 & e^{-i\rho(k)} \\ -e^{-i\rho(k)} & 0 \end{pmatrix}$, and the Berry connections satisfy $c_{22}(-k) = c_{33}(k) - \frac{\partial}{\partial k}\rho(k)$. These lead us to the charge polarization between spin-up bands $P_2$ as $P_2 = \int_0^\pi \frac{dk}{2\pi} C(k) - \frac{i}{2\pi}[\rho(\pi) - \rho(0)]$. Since $\rho(k) = i \log \vartheta_{23}(k)$, and $C(k) = tr(c(k))$, after a little algebra, we find

$$P_{tr} = i\int_0^\pi \frac{dk}{2\pi}\frac{\partial}{\partial k}\log(\det[\vartheta(k)]) - \frac{i}{\pi}\log\frac{\vartheta_{23}(\pi)}{\vartheta_{23}(0)} = \frac{i}{\pi}\cdot\frac{1}{2}\log\frac{\det[\vartheta(\pi)]}{\det[\vartheta(0)]} - \frac{i}{\pi}\log\frac{\vartheta_{23}(\pi)}{\vartheta_{23}(0)}.$$

$$P_{tr} = \frac{1}{i\pi}\log\left(\frac{\sqrt{\vartheta_{23}(0)^2}}{\vartheta_{23}(0)}\cdot\frac{\vartheta_{23}(\pi)}{\sqrt{\vartheta_{23}(\pi)^2}}\right). \qquad (7)$$

Obviously enough, the argument of the logarithm is +1 or −1. Furthermore, since $\log(-1) = i\pi$ one can see that $P_{tr}$ is 0 or 1 (mod 2). Physically, of course, the two values of $P_{tr}$ corresponds to two different polarization states which the system can take at t = 0 and t = T/2. The Bloch functions $|u^{(n)}(k,t)\rangle$ introduced above correspond to maps from the 2D phase space (k, t) to the Hilbert space. As in ref.**[31]**, the Hilbert space could be separated into two parts depending on the difference in $P_{tr}$ between t = 0 and t = T/2. This leads to introduction of a quantity $\nu$, specified only in mod 2, defined as ($P_{tr}$ (T/2)− $P_{tr}$ (0)). The Hilbert space is trivial if $\nu = 0$, while for $\nu = 1$ it is nontrivial (twisted). In our case, upon using Eq. (7), we obtain

$$(-1)^\nu = \prod_j \frac{\vartheta_{23}\left(k_{trim}^{(j)}\right)}{\sqrt{\vartheta_{23}\left(k_{trim}^{(j)}\right)^2}}. \qquad (8)$$

We have found above a single $k_{trim} = \left(2, \frac{\pi}{2}\right)$. A little calculation using Eq. (6) convinces us $\vartheta_{23}(k_{trim}) = -\vartheta_{32}(k_{trim})$. The square root of the square of the former is $\vartheta_{32}(k_{trim})$. Thus, as $\nu$ turns out to be 1, we find the Hilbert space to be twisted. The physical consequence of this nontriviality is the appearance of topologically-protected edge states. We now provide the outline of a different method below.

In the basis $(d_{k,\uparrow}^\dagger \ d_{k,\downarrow}^\dagger \ d_{k+Q,\uparrow}^\dagger \ d_{k+Q,\downarrow}^\dagger)^\dagger$, the starting Hamiltonian in Eq.(1) together with the Rashba coupling term in the DDW ordered phase ( for the ordering vector $\boldsymbol{Q}=(Q_1=\pm\pi, Q_2=\pm\pi)$) may be written in a compact form as $H(k_x, k_y) = (\varepsilon_k + C)\,\tau_3 \otimes s_0 + D_k\,\tau_1 \otimes s_0$ $+ (\Delta - g_1) \tau_3 \otimes s_1 + g_2 \tau_3 \otimes s_2$ where the additional terms are $\mathcal{E}$ and $\Delta$. The former corresponds to a random potential whose distribution is uniform in the interval $[-W/2, W/2]$ with a positive parameter W and $\Delta$ is the mass term. The matrices $\tau_k$ and $s_k$ (k = 1, 2, 3) are Pauli matrices for the orbital and the spin, respectively. Also, $g_1 = \alpha_0 \sin(k_y a)$ and $g_2 = \alpha_0 \sin(k_x a)$. The Hamiltonian $H(k_x, k_y)$ resembles the Bernevig-Hughes-Zhang(BHZ) model[32-34]. The Kane-Mele index $Z_2$ is calculated by us applying a novel method published by Katsura and Koma [35,36].The details is to be reported shortly in a separate communication. We show below that this method has a distinct advantages over other methods, viz. the index is determined here by the discrete spectrum of a supersymmetric structure endowed operator which enables one to estimate the index graphically.

As in ref. [35, 36], in two dimension (2D), let $P_F$ be the projection operator onto the states below the Fermi energy $E_F$ for infinite volume. We approximate this operator by the corresponding Fermi projection $P_F^{(\Lambda)}$ in the finite region $\Lambda$, where the matrix $P_F^{(\Lambda)} = \sum_{E_n \leq E_F} |u^{(n)}(k)\rangle\langle u^{(n)}(k)|$ and $|u^{(n)}(k)\rangle$ are eigenstates of the Hamiltonian $H(k_x, k_y)$ on $\Lambda$ corresponding to the eigenvalues $E_n$. We find that $E_n = (E_1, E_2)$ where $E_1 = -\sqrt{[\{(\varepsilon_k + W)^2 + D_k^2 + (g_1 - \Delta)^2 + g_2^2\} + \delta^2]}$, $E_2 = \sqrt{[\{(\varepsilon_k + W)^2 + D_k^2 + (g_1 - \Delta)^2 + g_2^2\} - \delta^2]}$, and $\delta^2 = \sqrt{[4(\varepsilon_k + W)^2((g_1 - \Delta)^2 + g_2^2) + D_k^4]}$. The corresponding eigenvectors, in the long wave-length limit, are

$$|u^{(1)}(k)\rangle = \begin{pmatrix} 1 \\ \psi^{(1)}(k) \\ 0 \\ 0 \end{pmatrix}, \text{ and } |u^{(2)}(k)\rangle = \begin{pmatrix} 1 \\ \psi^{(2)}(k) \\ 0 \\ 0 \end{pmatrix} \quad (9)$$

where $\psi^{(\beta)}(k) = \alpha_0(iak_x - ak_y + (\frac{\Delta}{\alpha_0}))/F_\beta$, $\beta = 1, 2$, and

$F_\beta =$

$[(E_\beta + \varepsilon_k + W)\{E_\beta^2 - (\varepsilon_k + W)^2 - D_k^2\} + (E_\beta - \varepsilon_k - W)\{(\alpha_0 a k_y - \Delta)^2 + (\alpha_0 a k_x)^2\}]/H_\beta^2$,

$$H_\beta^2 = [(E_\beta + \varepsilon_k + W)^2 - \{(\alpha_0 a k_y - \Delta)^2 + (\alpha_0 a k_x)^2\} + D_k^2]. \quad (10)$$

Equations (9) and (10), once again in the long wave-length limit, yield the 4× 4 matrix $P_F^{(\Lambda)} = \begin{pmatrix} M_{2\times2} & 0 \\ 0 & 0 \end{pmatrix}$ where $M_{2\times2} =$

$$\begin{pmatrix} 2 & (-\alpha_0 a k_y + \Delta) - i(\alpha_0 a k_x)/F \\ (-\alpha_0 a k_y + \Delta) + i(\alpha_0 a k_x)/F & [\{(\alpha_0 a k_y - \Delta)^2\} + \{(\alpha_0 a k_x)^2\}]/G^2 \end{pmatrix},$$

F = F₁F₂ / (F₁+F₂), and $G^2 = F_1^2 F_2^2 / (F_1^2 + F_2^2)$. We find the required eigenvalues of $P_F^{(\Lambda)}$ as $\nu_1 = 2$ and $\nu_2 \approx |\vartheta|^2 / F_2^2$ where $\vartheta = (-\alpha_0 a k_y + \Delta) - i(\alpha_0 a k_x)$. Whereas the eigenvalue $\nu_1 = 0$ specified in mod 2, a plot of $\nu_2$, as a function of the mass $\Delta$ and the strength W of disorder for the nodal and anti-nodal regions in Figure 3, show that $\nu_2 \approx 1$ when the random disorder is much less than one ( $W/t \ll 1$). However, when the disorder is moderately high, ($W/t \sim 0.1$) the system slides down to the state where $\nu_2 \approx 0$. The physical interpretation is that the former state (of the Bi2212 bilayer pseudo-gap(PG) state) corresponds to strong topological insulator state (as the Kane-Mele index $Z_2 = -1$), while the latter one corresponds to the weak topological insulator state of the system(as $Z_2 = 1$), provided we assume the PG state to be DDW ordered. The transition is occuring at zero temperature, hence this may be termed as the quantum phase transition(QPT). The yellow regions in Figure 3 are quantum critical region. A TRS breaking perturbations, such as the presence of the magnetic impurities, are expected to destroy topological surface state. Quite surprisingly, contrary to this expectation, we find that the surface state in Bi2212 is sensitive to the random disorder. This behavior is definitely not an isolated one. Kim et al.[37] have found that when the impurity concentration is moderate the surface state of SmB₆ (known to be a strong topological insulator **[38]**) can be altered even with non-magnetic impurities. In the next section, we discuss quantum anomalous Hall (QAH) effect under uniform exchange field when two bands close to Fermi energy are occupied.

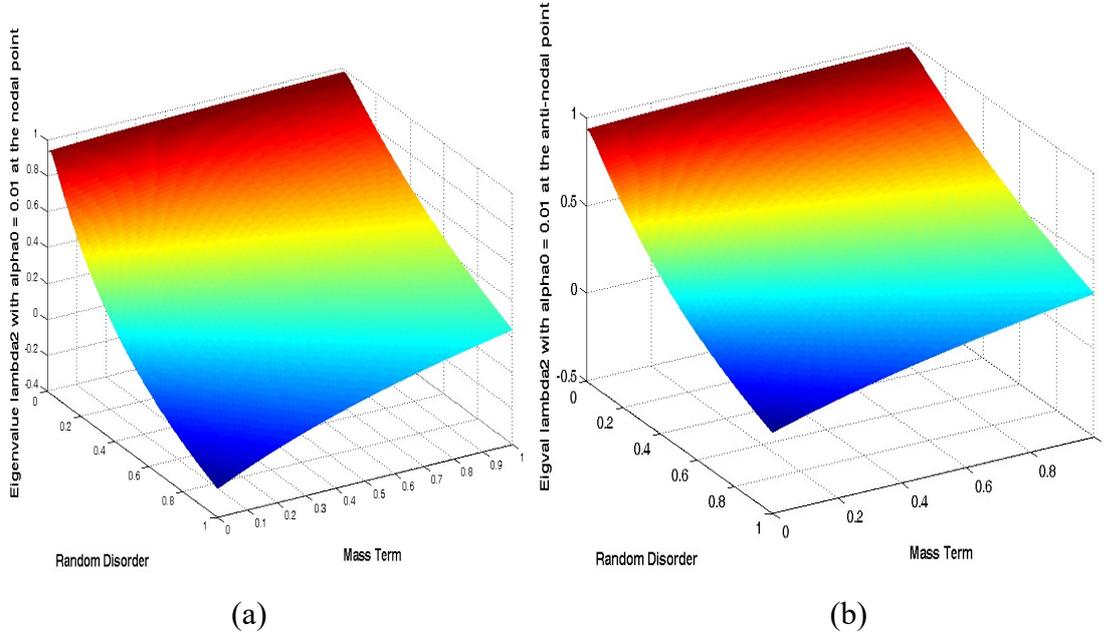

(a)          (b)

**Figure 3.** The plots of $\nu_2 \approx |\vartheta|^2 / F_2^2$ (where $\vartheta = (-\alpha_0 a k_y + \Delta) - i(\alpha_0 a k_x)$ and $F_2$ given by Eq. (10)) as a function of the mass $\Delta$ and the W of disorder for the (a) nodal($ak_x \sim \frac{\pi}{2}, ak_y \sim \frac{\pi}{2}$) and (b) anti-nodal ($ak_x \sim 0, ak_y \sim$)regions. The chemical potential $\mu \sim 0$. In figures (a), and (b), the common feature is that the eigenvalue $\nu_2$ is close to one when the disorder $W/t \ll 1$. This corresponds to the hot region or that corresponding to the strong topological insulator in the figures. However, for the moderately high disorder $W/t \sim 0.2$, the eigenvalue $\nu_2 \sim 0$, the system makes a transition to the ordinary insulating phase. The yellow region is quantum critical region. The other parameter values are, t =1 , t′/t = −0.28(hole-doping), t ″/t = 0.1, t ‴/t = 0.06, $t_p/t$ = 0.3, $t_{in}/t$ = 0.1, $\frac{\alpha_0}{t}$ = 0.5, $\Delta_0^{(PG)}$ (T)= 0.02, , $a_0$ = 0.40, $\mu$= −0.0, and $\chi_0$ = 0.0. The state dispersion is perfectly nested with the ordering wave vector **Q**=($Q_1=\pm\pi$, $Q_2= \pm\pi$).

## 3. Quantum Anomalous Hall Effect

A generic feature of all the QAH systems is that the bands close to the Fermi level correspond to those of chiral/ helical fermions (Dirac-like model). We shall use this model as a launch-pad for our discussion. Assuming that the system has magnetic impurities, we model the required interaction between an impurity moment and the itinerant (conduction) electrons in the system with coupling term $(J/t_1) \sum_m S_m \cdot s_m$, where $S_m$ is the $m$ th-site impurity spin, $s_m = \left(\frac{1}{2}\right) d^{\dagger}_{m\sigma} \sigma_z d_{m\sigma}$, $d_{m\sigma}$ is the fermion annihilation operator at site-$m$ and spin-state σ (=↑,↓) and $\sigma_z$ is the z-component of the Pauli matrices. We make the approximation of treating the impurity spins as classical vectors. The latter is valid for |**S**| >1. Absorbing the magnitude of the impurity spin into the coupling constant $J$ ($M = |J||S|/t_{d1}$) it follows that the exchange field term, in the $(d_{k,\uparrow} \quad d_{k,\downarrow} \quad d_{k+Q,\uparrow} \quad d_{k+Q,\downarrow})^T$ basis in the momentum space, appears as $\{\zeta_z \otimes M(\tau_0 + \tau_z)/2\}$, where $\tau_0$ and $\tau_z$, respectively, are the identity and the z-component of Pauli matrix for the pseudo-spin orbital indices. We thus obtain the dimensionless contribution $[M \sum_{k,\sigma} sgn(\sigma) d^{\dagger}_{k,\sigma} d_{k,\sigma}]$ to the momentum space Hamiltonian in Eq. (1) above. In other words, the exchange energy is given by the parameter M in first neighbor energy units.

We now turn our attention to the chiral DDW state. Our contention is that, with or without proximity to magnetic impurities, this state will display QAH effect. The weak-antilocalization (WAL) is induced by the chirality of the Dirac-like carries on the surface of such a system. A signature of WAL is usually obtained by calculating the magnetic field dependence of the conductivity correction described by the well-known Hikami-Larkin-Nagaoka (HLN) formula **[39]**. This is not a part of the investigation reported in this paper. Since (i) broken TRS, and (ii) strong RSOC are the requirements for the access of a QAH state which are very well fulfilled by the CDDW ordered ($d+id$) Hamiltonian in (1), we may surmise that the CDDW state is associated with QAHE. The plots of bands in $E_j(s,\sigma,k,k_z)$ in (4) in the d+id case belonging to the nodal and anti-nodal regions, as well as at Γ point are shown in Figures 4. The spin-momentum locking displayed here sets the stage for the further discussion of QAH effect. The numerical values of the parameters to be used in the calculation are t =1 , t′/t = −0.28(hole-doping), t ″/t = 0.1, t ‴/t = 0.06, $t_p/t$ = 0.3, $t_{in}/t$ = 0.1, $\frac{\alpha_0}{t} = 0.95, a_0 = 0.40, \mu = -0.0, \Delta_0^{(PG)}(T) = 0.02, \chi_0 = 0.02, M = 0, Q_1 = -0.7742$ π, and $Q_2 = -0.2258$.π. The state dispersion is imperfectly nested with the ordering wave vector **Q**=($Q_1$=±0.7742 π, $Q_2$= ±0.2258.π), **Q**=( $Q_1$= ±0.2258.π, $Q_2$=±0.7742 π ). We notice that in the nodal region spin-up holes are conducting, while around Γ point spin-up electrons are conducting. The reason for these observations is the corresponding bands are partially empty. At the anti-nodal region holes and electrons do not show conductance. Thus, the spin-momentum locking is evident, While the nodal region of the momentum space can give rise to a spin- up hole current, the region around the Γ point can give rise to spin-up electron current. The quasiparticle states of opposite spin are to be found in different parts of the Brillouin zone. On the experimental front, recently Vishwanath et al.[7] discovered that Bi2212, has a nontrivial spin texture with the spin-momentum locking. They used spin- and angle-resolved photoemission spectroscopic technique to unravel this fact. These authors alsodeveloped a model to show how this complex pattern could emerge in real and momentum spaces. The key feature is that the layered structure of Bi2212 allows for a spin-momentum locking (SML) in one Cu-O layer of the unit cell that is matched by the opposite spin texture in the other Cu-O layer of the unit cell through the Γ point encirclement in the momentum space representation (see Figure 1 and ref.7). This suggests the presence of a strong spin-orbit coupling in Bi2212 like a topological insulator. We shall discuss briefly the encirclement feature below.

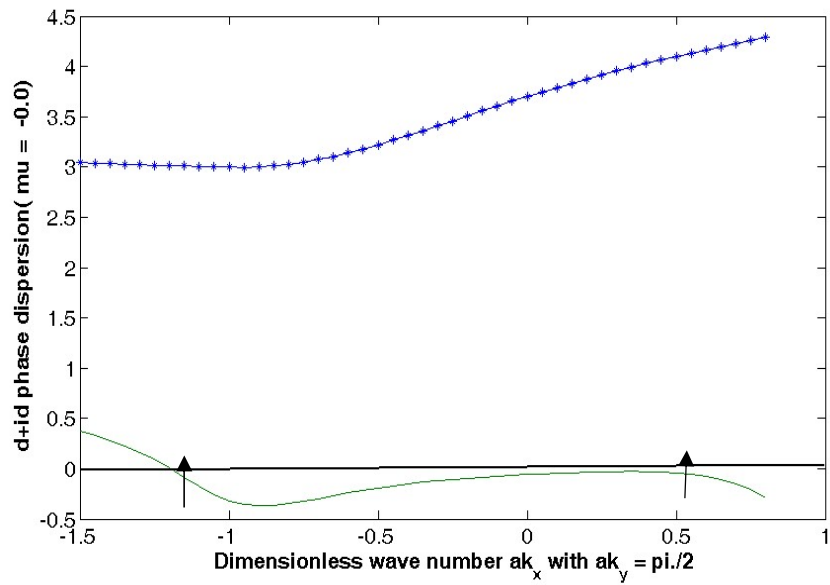

(a)

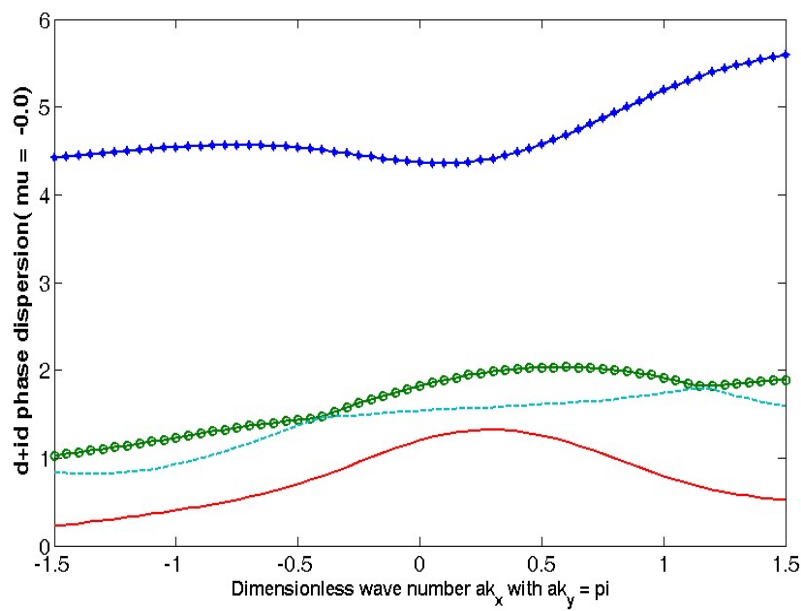

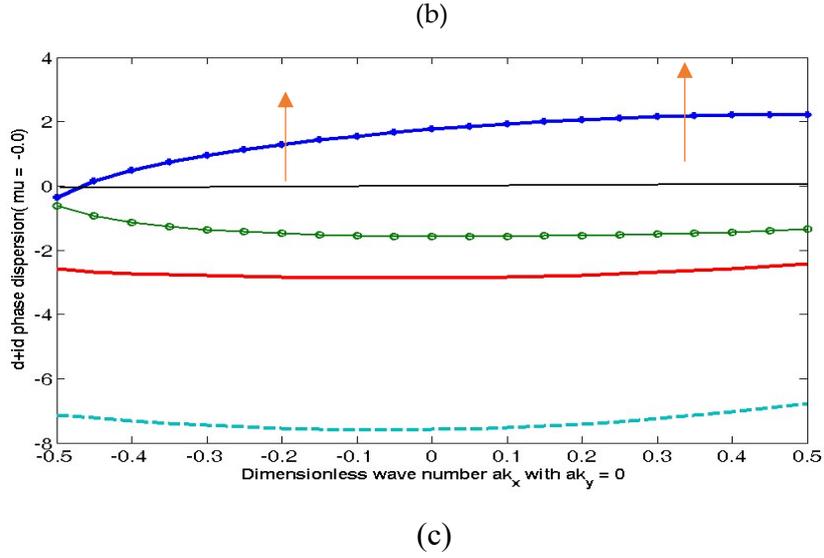

(b)

(c)

**Figure 4.** The plots of energy bands $E_j(s,\sigma,k,k_z)$ in (4) in the d+id (QAH) case belonging to the nodal and anti-nodal regions, as well as at Γ point are shown in Figures 4. The numerical values of the parameters to be used in the calculation are t =1 , t'/t = −0.28(hole-doping), t ''/t = 0.1, t '''/t = 0.06, $t_p$/t = 0.3, $t_{in}$/t = 0.1, $\frac{a_0}{t}$ = 0.95, $a_0$ = 0.40, μ= −0.0, $\chi_0$= 0.02, M = 0, $Q_1$ = -0.7742 π, $Q_2$ = -0.2258.π, and $\Delta_0^{(PG)}$ (T)= 0.02. The state dispersion is imperfectly nested. (a)In the nodal region spin-up holes are conducting, (b) while at the anti-nodal region holes and electrons do not show conductance. (c) At the region around Γ point spin-up electrons are conducting. The reason for these observations is the corresponding bands are partially empty.

The spin texture **s**(n,**k**) is defined as the expectation value of a vector operator $S_j = I_{2 \times 2} \otimes \sigma_j$ where $\sigma_j$ are Pauli matrices on a two dimensional **k**-grid and $\otimes$ stands for the tensor product. At **k** for the state n (or nth band) it is defined as an expectation value $s_j(n,\mathbf{k}) = \langle S_j \rangle^{(n)} = \langle n|S_j|n\rangle$. Obviously enough, to calculate this we need eigenvectors of the Hamiltonian matrix given by Eq.(1) for the eigenvalues

$$E_1 = \sqrt{\frac{\eta_0(k)}{2} + \varepsilon_k^U - \left(b_0(k) - \left(\frac{\eta_0(k)}{2}\right)\right) + c_0(k)\sqrt{\frac{2}{\eta_0(k)}}}^{\frac{1}{2}} \pm \varepsilon_{k_z}(k),$$

$$E_2 = \sqrt{\frac{\eta_0(k)}{2} + \varepsilon_k^U - \left(b_0(k) - \left(\frac{\eta_0(k)}{2}\right)\right) + c_0(k)\sqrt{\frac{2}{\eta_0(k)}}}^{\frac{1}{2}} \pm \varepsilon_{k_z}(k), \qquad (9)$$

These eigenvectors will be different from what is given in Eq. (5) as the interplanar tunneling term $\varepsilon_{k_z}(k)$ has been inducted in the eigenvalues ( $E_j(s,\sigma,k,k_z) = \epsilon_j(s,\sigma,k) \pm \varepsilon_{k_z}(k)$). In fact, the replacement $\epsilon_j(s,\sigma,k) \to E_j(s,\sigma,k,k_z)$ in Eq. (5) will yield the eigenvectors sought for. We have calculated expectation value $s_j(n,\mathbf{k})$ with these vectors. The 2D plots of the expectation values shown in Figure 5 have been drawn using the numerical values of the parameters as t =1 , t'/t = −0.28(hole-doping), t ''/t = 0.1, t '''/t = 0.06, $t_p$/t = 0.3, $t_{in}$/t = 0.1, $\frac{\alpha_0}{t} = 0.95, a_0 = 0.40, \mu = -0.0, \Delta_0^{(PG)}(T) = 0.02, \chi_0 = 0.02, Q_1 = 0.7742\,\pi,$ and $Q_2 = 0.2258\,\pi.$ $\frac{\alpha_0}{t} = 0.95, t_{in}/t = 0.1.$ We find that unfailingly, near the nodal point, the expectation values $s_j(n,\mathbf{k})$ show peaks when the numerical values of the parameters are $t = 1$, $t_p/t = 0.3$, $t_{in}/t = 0.1$, $\frac{\alpha_0}{t} = 0.1, 0.2, 0.3, a_0 = 0.40,$ and $\mu = -0.0$. As shown in the 3D plot of $s_z(n,\mathbf{k})$ in Figure 5(f) and (g), the spike in this quantity is broadly confined to the nodal region. The Figures (a), (b), and (c) show that, when the interlayer tunneling $t_{in}$ is kept fixed, the quantity $s_z(n,\mathbf{k})$ is an increasing function of the RSOC coefficient $\frac{\alpha_0}{t}$. On the other hand, (a), (d), and (e) show that, for a fixed $\frac{\alpha_0}{t}$, $s_z(n,\mathbf{k})$ is a decreasing function of the interlayer tunneling $t_{in}$. The origin of this tunability is as follows: A consequence of the so-called structural inversion symmetry is the Rashba spin-orbit coupling (RSOC). This can be tuned by gate voltage or doping. As shown in Figure 5(g) (where we find that the texture almost disappears when RSOC tends towards zero), we notice that the non-trivial spin texture in k-space (spike in the nodal region) is a consequence of RSOC. The texture, therefore, should be tuneable by electric field ($\alpha_0$ is electric field dependent) and intercalation (conjecturally, this will affect the interlayer tunneling). This is precisely we find above. Thus, at $k_z$ = a rational number plane (and in a large region of the Brillouin zone), we find that the tuneable spin textures demonstrates a momentum-selective spin polarization, which has a strong implication to spintronic applications.

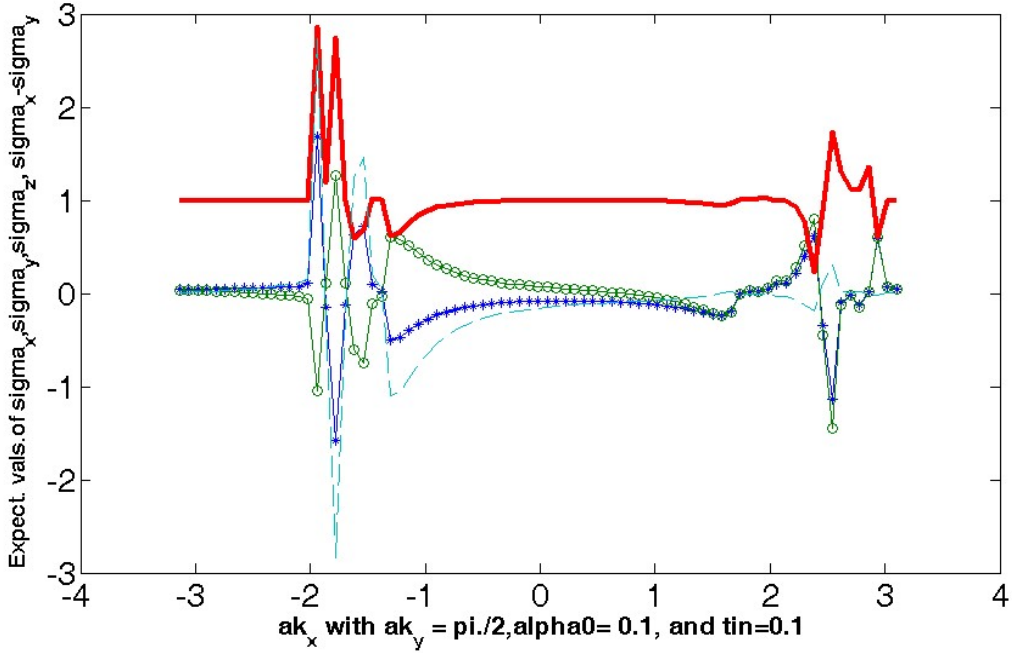

(a)

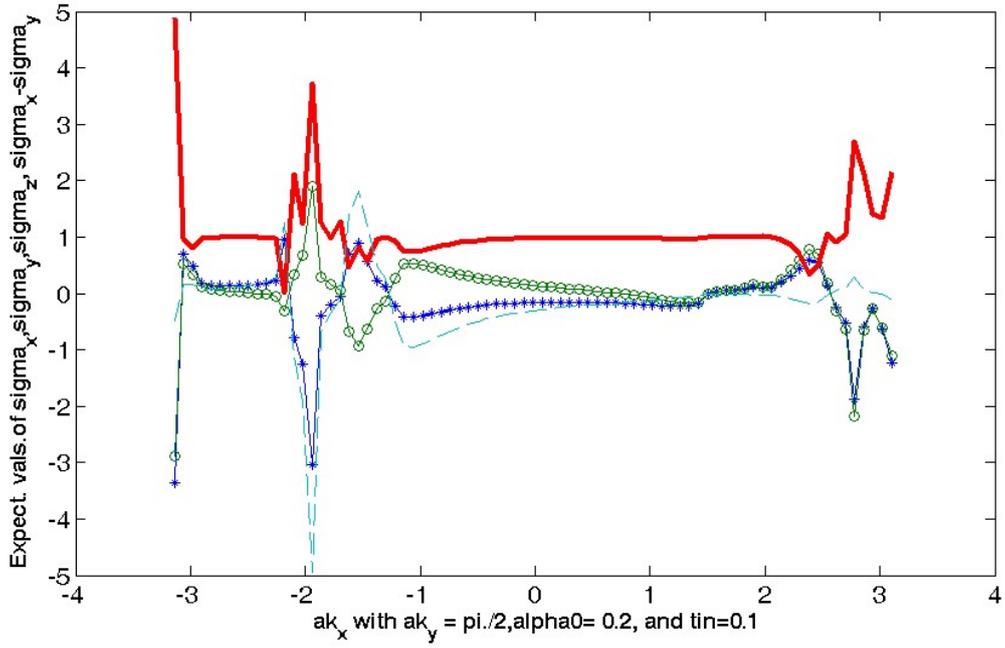

(b)

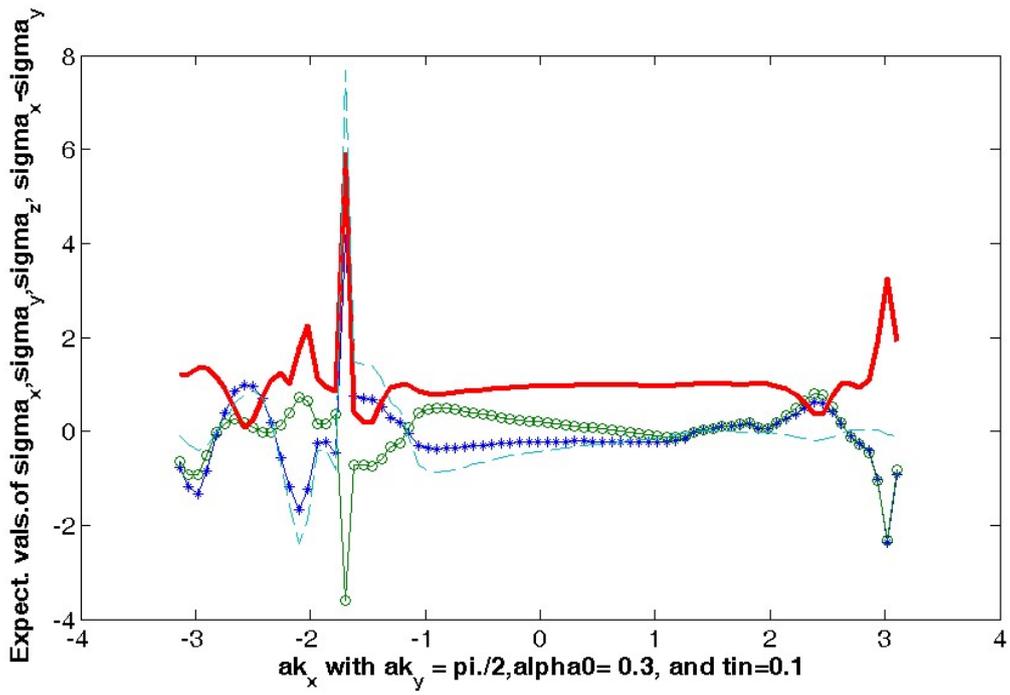

(c)

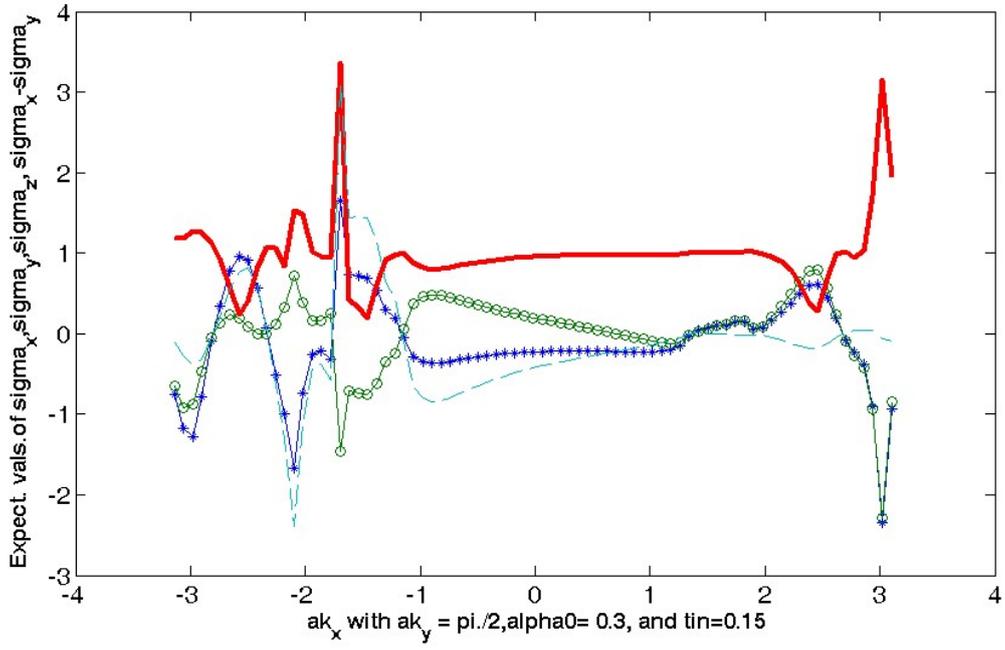

(d)

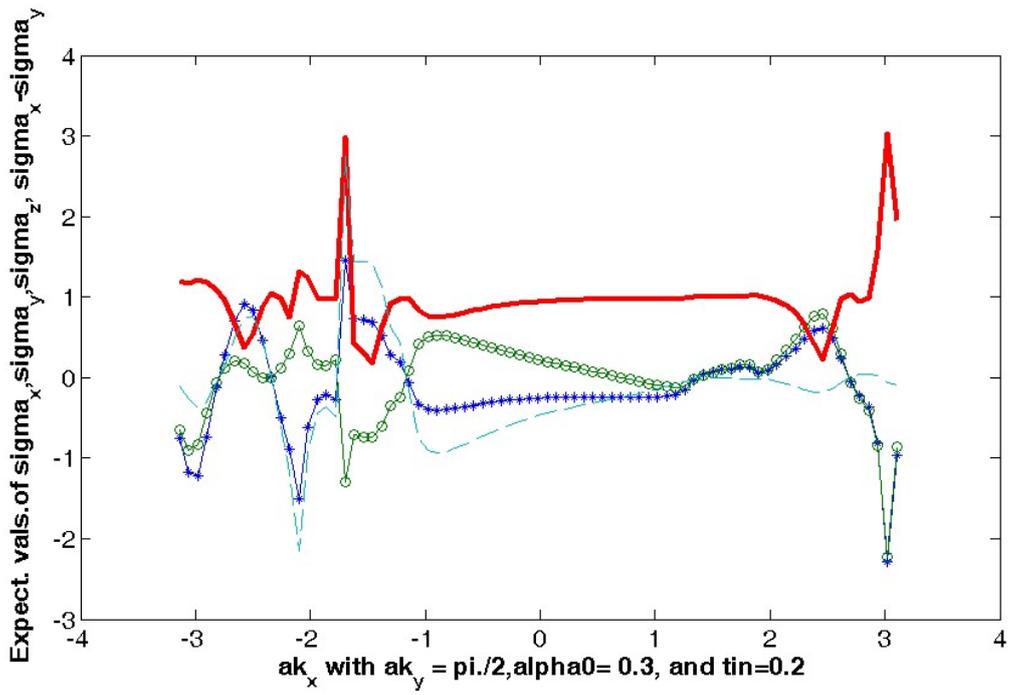

(e)

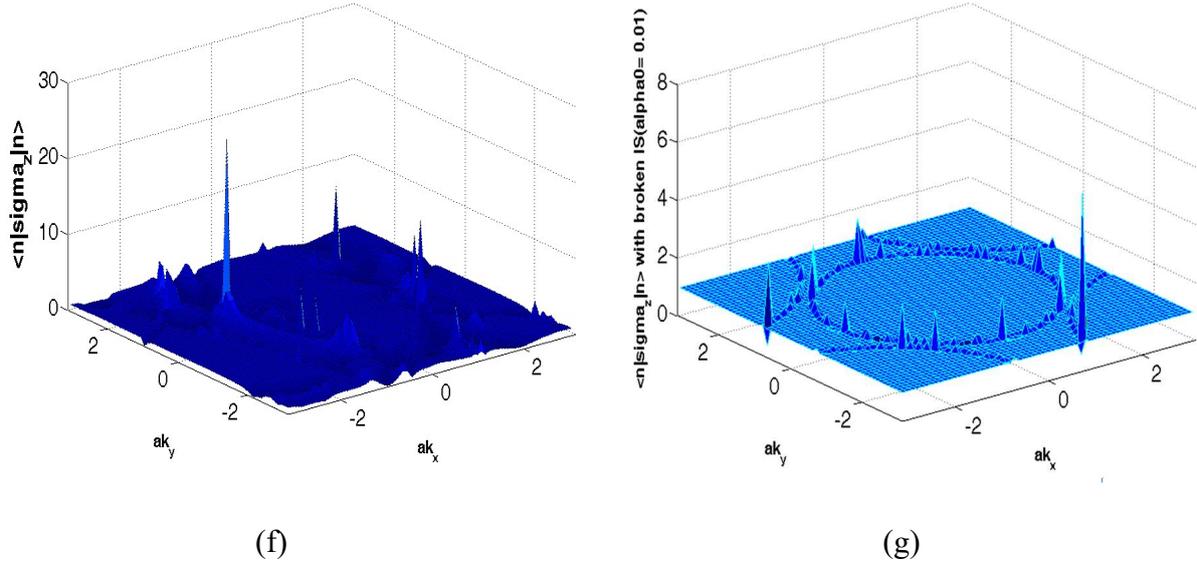

(f)                                         (g)

**Figure 5. (a)-(f)** The plots of the spin texture (the expectation value of the spin operator $S_z$) as a function of momentum. The numerical values of the parameters as t =1 , t′/t = −0.28(hole-doping), t ″/t = 0.1, t ‴/t = 0.06, $t_p/t$ = 0.3, $t_{in}/t$ = 0.1, $\frac{\alpha_0}{t}$ = 0.95, $a_0$ = 0.40, μ= −0.0, $\Delta_0^{(PG)}(T)$= 0.02, $\chi_0$ = 0.02, $Q_1$ = 0.7742 π, and $Q_2$ = 0.2258.π. $\frac{\alpha_0}{t}$ = 0.95, $t_{in}/t$ = 0.1. **(g)** The 3D plot of the spin texture showing disappearing spikes in the nodal region. The numerical values of the parameters used are t =1 , t′/t = −0.28(hole-doping), t ″/t = 0.10, t ‴/t = 0.06, $t_b/t$ = 0.30, $t_z/t$ = 0.10, $\frac{\alpha_0}{t}$ = 0.01, $\frac{\chi_0}{t}$=0.44, $\frac{\Delta_0^{PG}}{t}$ = 0.20 and $a_0$ = 0.40.

We have considered a Bloch Hamiltonian *H(k)* to describe our system. The Hamiltonian we start with is nearly the same as that put forward by Das Sarma et al. **[4-6]** when we do not consider RSOC. Their Hamiltonian is inversion symmetry protected and time reversal symmetry broken. The inversion symmetry is protected as long as the ordering wave vector **Q**=(±π,±π). For **Q** ≠(±π,±π), the inversion symmetry protection is lost even in the absence of the Rashba term. Suppose the eigenvectors corresponding to the energy eigenvalues of *H(k)* are denoted by $u^{(n)}(k)$, where *n* is a band index. Suppose $u^{(n)}(k)$ is a Bloch state eigenwavefunction. The Berry curvature (BC) is defined as

$$\Omega_{xy}^{(n)}(k) = i \left\langle \partial_{k_x} u^{(n)}(k) \big| \partial_{k_y} u^{(n)}(k) \right\rangle - i \left\langle \partial_{k_y} u^{(n)}(k) \big| \partial_{k_x} u^{(n)}(k) \right\rangle \quad (11)$$

Two important symmetries one needs to consider here are time reversal symmetry (TRS) and inversion symmetry (IS). For the inversion symmetry(IS), $\Omega_{xy}^{(n)}(k) = \Omega_{xy}^{(n)}(-k)$ and, for the time reversal symmetry (TRS), $\Omega_{xy}^{(n)}(k) = - \Omega_{xy}^{(n)}(-k)$. Thus, for a system which is both TRS and IS compliant, $\Omega_{xy}^{(n)}(k) \equiv 0$. This means that in order to study the possible quantum anomalous Hall (QAH) effect starting from the present DDW ordered TRS (IS) compliant (noncompliant) system of Bi2212 bilayer we need to have broken TRS, i.e. non-zero Berry curvature (BC). A non-zero BC is very much required to obtain anomalous Hall conductivity $(\sigma_{xy})$ and to show that $\sigma_{xy}$ is quantized in the case of an insulator. The Dirac model approach though has its limitations, it is nevertheless useful in understanding the topological properties of realistic materials. Since the intrinsic QAH effect can be expressed in terms of BC, it is therefore an intrinsic quantum-mechanical property of a perfect crystal. An extrinsic mechanism, skew scattering from disorder, tends to dominate the QAH effect in highly conductive ferromagnets. Since bands close to the Fermi level correspond to those of chiral/helical fermions (Dirac-like model), as already stated, we consider the following (derivable from(1) assuming long wavelength limit): $\epsilon_1 \approx \varepsilon_k - \sqrt{M^2 + \alpha_0^2(x^2 + y^2) + O(\varsigma^2)}$ and $\epsilon_2 \approx -\varepsilon_k + \sqrt{M^2 + \alpha_0^2(x_1^2 + y_1^2) + O(\varsigma^2)}$ where $x = ak_x, y = ak_y, x_1 = x + Q_1$, $y_1 = y + Q_2$, and $\varsigma = (\chi_0/2) xy + (\frac{i}{4})\Delta_0^{(PG)}(T)(x^2 + y^2)$. Upon using the formula in Eq.(11) we find the BC ($\Omega_{xy}$) as $[2((B_1B_2)-(A_1A_2)+ 2((B_{21}B_{22})-(A_{21}A_{22}))]$. Here

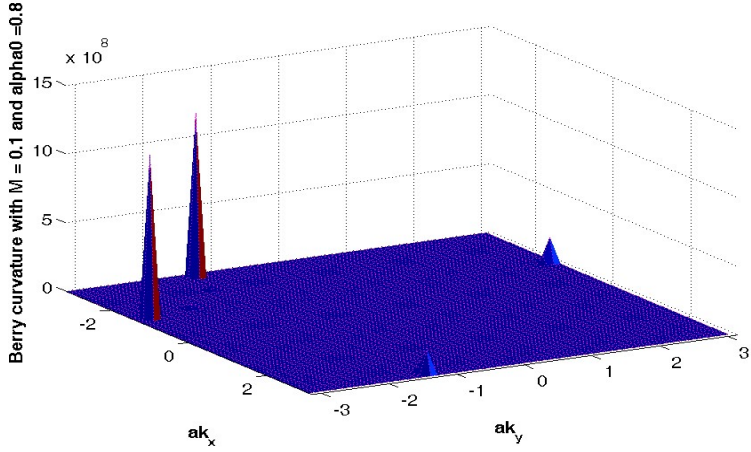

(a)

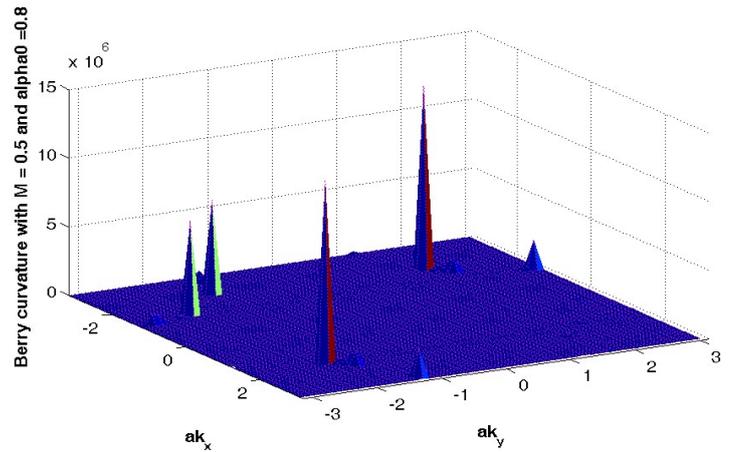

(b)

**Figure 6.** The plots of BC as a function of $(ak_x, ak_y)$. The chemical potential $\mu \sim 0$. In figures (a), and (b), the common feature is the spikes in certain regions of the Brilloin zone close to the nodal points. (a) For the low exchange field value $M/t = 0.1$, there are fewer spikes. (b) For the moderately high value $M/t = 0.5$, there

are larger peaks. The other parameter values are, t =1 , t′/t = −0.28(hole-doping), t″/t = 0.1, t‴/t = 0.06, $t_p$/t = 0.3, $t_m$/t = 0.1, $\frac{\alpha_0}{t}$ = 0.8, $\Delta_0^{(PG)}$ (T)= 0.02,, $a_0$ = 0.40, μ= −0.0, and $\chi_0$ = 0.02.

$$A_1 = \alpha_0\{((\epsilon_1 - \varepsilon_k) + M) - x(\partial \epsilon_1/\partial x)\}/((\epsilon_1 - \varepsilon_k) + M)^2 \; ; A_2 = A_1 (x \to y),$$
$$B_1 = \alpha_0\{y(\partial \epsilon_1/\partial x)\}/((\epsilon_1 - \varepsilon_k) + M)^2 \; ; B_2 = B_1 (x \to y),$$
$$A_{21} = A_1(\epsilon_1 \to \epsilon_2, x \to x_1, y \to y_1), \; A_{22} = A_2(\epsilon_1 \to \epsilon_2, x \to x_1, y \to y_1).$$
$$B_{21} = B_1(\epsilon_1 \to \epsilon_2, x \to x_1, y \to y_1), \; B_{22} = B_2(\epsilon_1 \to \epsilon_2, x \to x_1, y \to y_1). \quad (12)$$

The plots of BC as a function of $(ak_x, ak_y)$ with the chemical potential μ ~0 are shown in Figure 6. In figures 6(a), and 6(b), the common feature is the spikes in certain regions of the Brilloin zone close to the nodal points. There are fewer spikes in (a), whereas in (b) there are larger number of peaks. While the exchange field value $M/t$ = 0.1 in (a), in (b) the value $M/t$ = 0.5 is moderately high value. As we have not dealt with a large number of cases, the tentative conclusion is the proliferation of spikes with the increase in the exchange field. The other parameter values are, *t =1 , t′/t = −0.28*(hole-doping), *t″/t = 0.1, t‴/t = 0.06, $t_p$/t = 0.3, $t_m$/t = 0.1,* $\frac{\alpha_0}{t}$ = 0.8, $\Delta_0^{(PG)}$ *(T)= 0.02,, $a_0$* = 0.40, *μ*= −0.0, and $\chi_0$ = 0.02. The famous TKNN formula states that the Hall conductivity $\sigma_{xy}$ in a band insulator is related to the Chern number C by $\sigma_{xy}$ = $(e^2/2h)$ C and the Chern number C, which is an integer, is given by C = 2 $\iint_{BZ} \Sigma_n \Omega_{xy}^{(n)}(k) \frac{d^2k}{(2\pi)^2}$. Here BZ denotes an integral over the entire Brillouin zone (BZ), and the sum is over occupied bands. The Chern numbers are particularly easy to evaluate in two-band models. To obtain the Hall conductivity, k-integration is necessary. For the integration purpose, we first divide the BZ into finite number of rectangular patches. We next determine the numerical values corresponding to each of these patches of the momentum-dependent density and sum these values. The sum is then divided by the number of patches. We have generated these values through a surface plot. We find, for example, C = 0.7488 for $M/t$ = 0.5. There is a caveat which concerns the validity of the calculated result. The Hall conductivity $\sigma_{xy}$ cannot be determined as such from the 2D Dirac model since the former requires an integral over the whole BZ, as we have seen above. The integral is outside the Dirac model's range of validity. To circumvent the problem we can choose a momentum space cut-off small compared to the size of BZ and large enough to capture nearly all the Berry curvature integral. This will be within the range of validity of the 2D Dirac model. In this limit, the Hall conductivity $\sigma_{xy} \to C(e^2/2h)$ where $C \to$ an integer. The sign of the Hall contribution is determined by the relative signs of BCs. In the case when the system is insulator, C= $(1/2\pi) \oint \Sigma_n d\mathbf{k}. \mathbf{A_n}$ where the closed loop corresponds to the encirclement of the Brillouin zone boundary. Here $\nabla_k \times A_n(\mathbf{k}) = \Omega_{xy}^{(n)}(k)$ is the Berry curvature (BC).

The anomalous Hall conductivity is then quantized. The reason is because of the single-valued nature of the wave function, its change in the phase factor after encircling the Brillouin zone boundary can only be an integer multiple of $2\pi$ or $2\pi m$ ( the Berry connection is not a single-valued function and so the TKNN invariant becomes nonzero, i.e. the system becomes topological) which means C can only take an integer value, and hence $\sigma_{xy}$ is quantized to integer multiples of $e^2/h$. The integer is called TKNN invariant, and it plays the role of the topological invariant of the quantum Hall system.

## 4. Discussion and conclusion

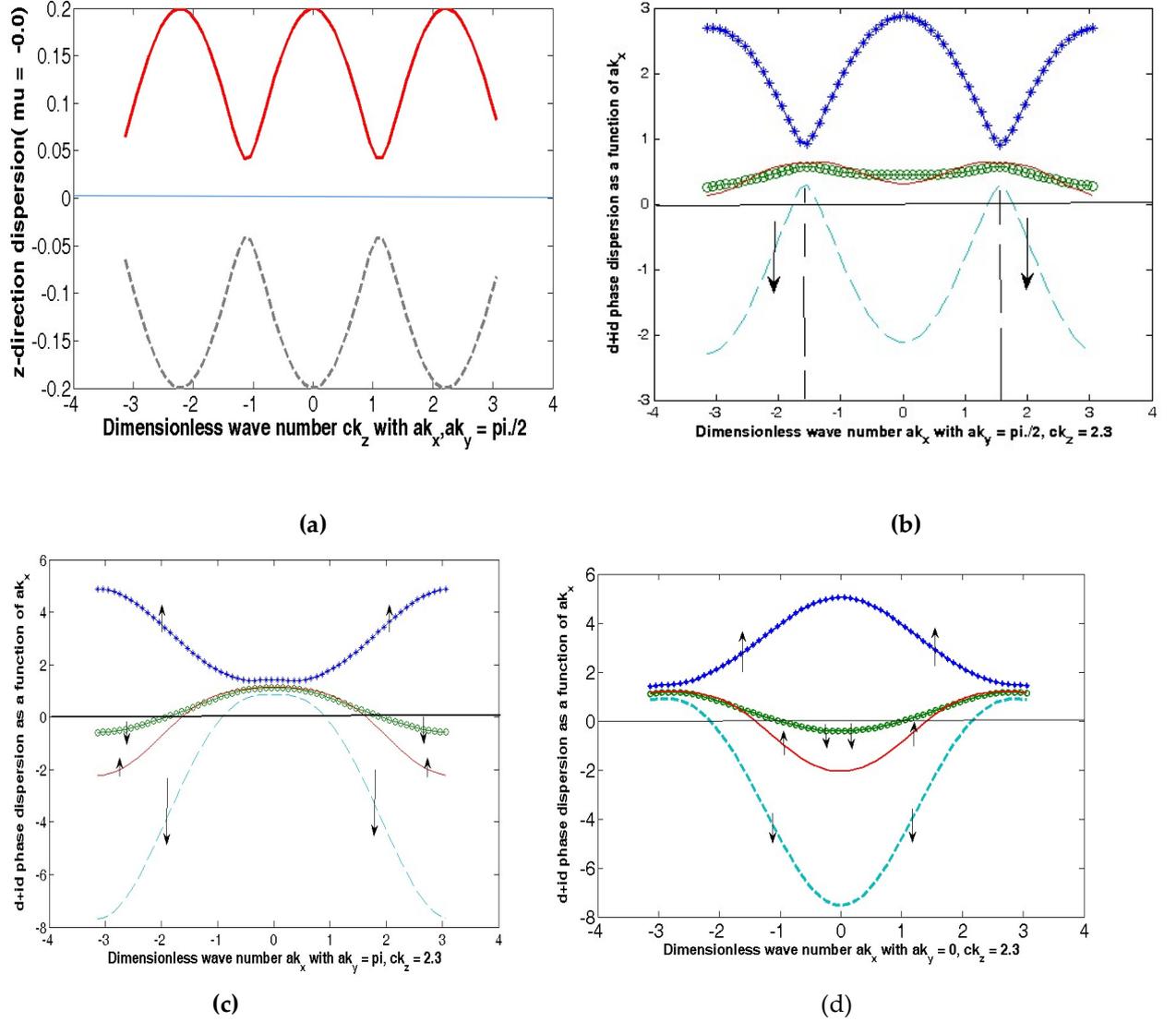

**Figure 7.** The 2D-plots of quasiparticle excitation bands in Eq. (2) as function of dimensionless momentum $ak_j$ in the absence of the Rashba coupling. The numerical values of the parameters used are t =1 , t'/t = −0.28(hole-doping), t''/t = 0.1, t'''/t = 0.06, $t_b$/t = 0.3, $t_z$/t = 0.1, $\frac{\alpha_0}{t} = 0$, $\frac{\alpha_1}{t} = 0$, $\frac{\chi_0}{t}$=0.44, $\frac{\Delta_0^{PG}}{t} = 0.2$, and $a_0 = 0.4$.

We have added in the Rashba term by hand. Here we provide a sufficiently strong supporting argument from band structure calculation is as follows: In fact, there is lack of the evidence of the spin-momentum locking when the Rashba coupling is totally absent. This necessitates its inclusion. To justify this statement, we show the plots of the energy eigen-values in Eq.(1) of the manuscript(MS) in the nodal and anti-nodal regions are shown in Figures 7 *(a) − 7(d)* above in the absence of the Rashba coupling. The numerical values of the parameters used are t =1 , t′/t = −0.28(hole-doping), t ″/t = 0.1, t ‴/t = 0.06, t_b/t = 0.3, t_z/t = 0.1, $\frac{\alpha_0}{t} = 0$, $\frac{\alpha_1}{t} = 0$, $\frac{\chi_0}{t} = 0.44$, $\frac{\Delta_0^{PG}}{t} = 0.2$, and $a_0 = 0.4$. In Figure (a) we have a plot of quasi-particle excitation (QP) spectrum in z-direction given by $\varepsilon_{k_z}(k) = - Y_z(k, s_z(c/2))[(s_x(a) - s_y(a))^2 /4 + a_0]$ given above as function of dimensionless momentum $ak_z$ at the nodal point. In Figures (b), (c), and (d) we have plots of the excitation spectrum in the CDDW state as a function of $ak_x$ with $ak_z = 2.3$. In Figure (b), the plot is for $ak_y = \frac{\pi}{2}$. We find that only the spin-down valence band is partially empty. In Figure (c) and (d), respectively, the plot is for $ak_y = \pi$ and $ak_y = 0$. We find that only the spin-up conduction band is completely full whereas the rest of them partially empty. Thus, while the nodal region of the momentum space can give rise to a spin-down hole current, no other region can give rise to spin- polarization. We conclude that the spin-momentum locking is not evident in the absence of the Rashba couplinng in Bi2212 bilayer system. This is a very strong argument in favor of the inclusion of the Rashba coupling for the problem in hand.

Vishwanath et al.[7] discovered that Bi2212, has a nontrivial spin texture with the spin-momentum locking. They used spin- and angle-resolved photoemission spectroscopic (SARPES) technology to unravel this fact. They also developed a model to show how this complex pattern could emerge in real and momentum spaces. The key feature is that the layered structure of Bi2212 allows for a spin-momentum locking in one Cu-O layer of the unit cell that is matched by the opposite spin texture in the other Cu-O layer of the unit cell, as shown in Figure1, through the Γ point encirclement in momentum space representation. This type of spin-momentum ordering is hidden from other experimental techniques except SARPES [7]. A similar figure is displayed also in Figure 5(f) here. As shwn in the plot of the texture **s**_z(n,**k**) in this figure, the spike in **s**_z(n,**k**) is broadly confined to the nodal region and there is encirclement of Γ point.

The Rashba spin-orbit coupling (RSOC), which can be tuned by gate voltage or doping, is a consequence of the so-called structural inversion symmetry. On the other hand, SOC linked to the bulk inversion symmetry is referred to as the Dresselhaus coupling (D-SOC). Here both k-linear and k-cubed terms are involved[40]. The linear term is given by : $\beta_0(\sigma_x \sin(k_x a) - \sigma_y \sin(k_y a))$ where $\beta_0$ is the Dresselhauss spin-orbit coefficient [40,41]. The cubic term, on the other hand, is $\gamma_0(\sigma_x (k_y a)^2 (k_x a) - \sigma_y (k_x a)^2 (k_y a))$ where $\gamma_0$ is the Dresselhauss spin-orbit coefficient. It is well-known **[40,41]** that the coexistence of Rashba and Dresselhaus contributions is possible in a two-dimensional spin-orbit coupled system without inversion symmetry. Therefore we shall include the D-SOC term in the QSH Hamiltonian in our next communication. We intend to observs the effect of cubic term in spin-momentum locking, QSH and QAH effects. In fact, there remains much to be understood and explored in this field. The model we wish to consider might be realized with cold atoms. The reason being the Rashba and Dresselhaus spin-orbit coupling can be realized in cold atoms **[42-45]** in an optical lattice experimentally.

The important question here is "what is the necessity of breaking inversion symmetry (IS)". In their seminal work, Zhang et al. **[46]** have demonstrated that the lack of the local inversion symmetry at atomic sites leads to hidden spin polarization completely determined by the site-dependent orbital angular momentum even in centrosymmetric crystals. This was an outcome of the first-principles calculations by the authors. Usually, the reason behind this is the orbital magnetization being more important than the spin magnetization, i.e. the spin–orbit coupling (SOC) is weak, such as in a transition metal dichalcogenide (TMD) $MoS_2$. Additionally, Baidya et al. **[47]** had also explored the important role of orbital polarization in QAH phases. A comparison with DFT calculation led these authors to the conclusion that effect of such terms are smaller.

In summary, we have possibly for the first time found that a DDW ordered **[1-3]** PG state model of Bi2212 bilayer, which involves the Rashba spin-orbit coupling, can display QSH effect. This model describes the low energy physics of a class of helical fermionic liquid in our spin-orbit system. We also predict the existence of QAH effect in the context of an extension of the DDW model, viz. chiral DDW ordered state **[4-6]** with magnetic impurities. This spin-orbit coupled systems have potential technological applications in spintronics. There are plenty of future challenges and opportunities of this rapidly evolving area.